\documentclass[twocolumn, aps,prl]{revtex4}
\usepackage{graphicx}
\usepackage{amsfonts}
\usepackage{amsmath}
\usepackage[varg]{txfonts}
\usepackage{bm}
\usepackage{mathrsfs}
\newcommand{\dg}{\tfrac{\gamma_R\gamma_D}{2}}
\newcommand{\E}{\tilde{E}}
\newcommand{\D}{\mathscr{D}}
\newcommand{\gp}{\gamma_+}
\newcommand{\gm}{\gamma_-}
\newcommand{\pspace}{(\gamma_R,\gamma_D)}

\graphicspath{{./graphics/}}

\begin{document}

\title{Bulk Spin-Hall Effect}
\author{Brandon Anderson, Tudor D. Stanescu, and Victor Galitski}
\affiliation{Condensed Matter Theory Center and
Joint Quantum Institute, Department of Physics, University of
Maryland, College Park, MD 20742-4111}

\begin{abstract}
We show that a two-dimensional spin-orbit-coupled system in the presence of a charge/spin-density wave with a wave-vector perpendicular to an applied electric field supports bulk manifestations of the direct/inverse spin-Hall effect. We develop a theory of this phenomenon in the framework of the spin diffusion equation formalism and show that, due to the inhomogeneity created by a spin-grating, an anomalous bulk charge-density wave is induced away from sample boundaries.
The optimal conditions for the observation of the effect are determined. The main experimental manifestation of the bulk spin-Hall effect, the induced charge/spin-density-wave, is characterized by a $\pi/2$-phase shift relative to the initial non-homogeneous spin/charge-polarization profile and has a non-monotonic time-varying amplitude.
\end{abstract}


\maketitle

Electron spin transport in semiconductors has recently evolved into a subject of intense research as key element of the rapidly developing field of spintronics~\cite{spintronics1,spintronics2}. One of the main challenges is to generate spin polarization and to transport spin in nonmagnetic materials using electric fields, by taking advantage of the coupling between spin and orbital degrees of freedom~\cite{Dress,Rash}. Of particular importance in this respect are a family of anomalous transverse transport phenomena, such as the spin Hall effect~\cite{SCNSJM,MNZ}, which has received recently tremendous attention. The experimental manifestation of the effect is the appearance of spin accumulation near the edges of the sample\cite{Kato,WKSJ,Sih}, if an electric current is driven through a system with either intrinsic or extrinsic (i.e., impurity-induced) spin-orbit coupling~\cite{EHR}. Hence, the canonical spin-Hall effect is, at least from an experimental perspective, an edge phenomenon whose magnitude  depends strongly on the specific boundary conditions~\cite{GBS,SurfS}. The role of the edge is to create a strong inhomogeneity where the experimentally observable spin density can accumulate. However, it is possible to create large length scale inhomogeneities in a controlled way, for example by generating a modulated charge or spin distribution or a spatially varying spin-orbit coupling. By analogy with the canonical  edge spin-Hall effect, the externally generated charge/spin-densities  would effectively create multiple boundaries in the bulk, thus opening the possibility of realizing bulk manifestations of the direct/inverse spin-Hall effect and allowing for a controlled study of this phenomenon.

\begin{figure}[tbp]
\begin{center}
\includegraphics[width=0.48\textwidth]{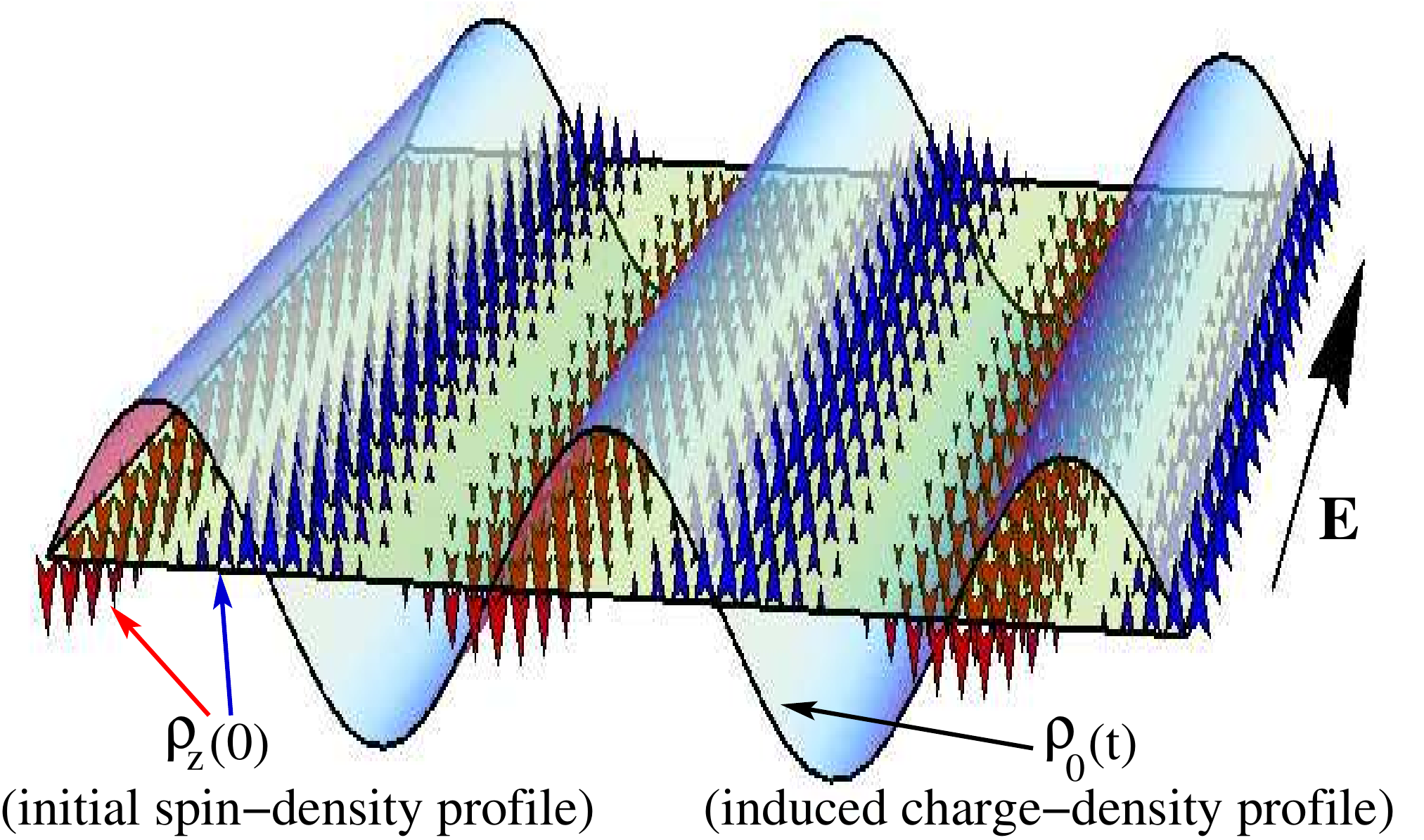}
\end{center}
\caption{(Color online) Charge-density profile induced by the relaxation of a spin-density wave in the presence of a uniform electric field. The initial spin density corresponds to a sinusoidal wave with wave-vector ${\bf q}$ of the out-of-plane $S_z$ component, as symbolized by the blue (spin up) and red (spin-down) arrows. The in-plane electric field is oriented perpendicular to ${\bf q}$. Notice the $\frac{\pi}{2}$ shift of the induced charge-density profile relative to the spin-density wave. \vspace*{-0.15in}} \label{Fig1}
\end{figure}

An effective way of producing bulk manifestations of the spin-Hall effect, suggested by the recent work of Koralek et al.~\cite{orenstein}, is to use the transient spin grating (TSG) technique~\cite{TSG1,TSG2,weber} to generate and monitor time dependent spin and charge profiles. Within the TSG method, a sinusoidal spin-polarization wave is generated by two interfering non-collinear laser beams with orthogonal linear polarization. This induces a modulation in the index of refraction, which can be measured at subsequent times by the diffraction of a probe pulse. In the presence of an external electric field oriented perpendicular to the spin-polarization wave-vector, a charge density with the same wave-vector is expected to develop (see Fig. \ref{Fig1}). Alternatively, in a spin-orbit interacting system a charge density wave is expected to induce spin modulation under the action of an external electric field.

In this Letter, we develop a theory of the bulk spin-Hall effect in the diffusion limit, in the presence of Rashba~\cite{Rash} and (linear and cubic) Dresselhaus~\cite{Dress}  spin-orbit interactions. We focus on the time evolution of a charge density profile induced by an optically generated spin-polarization wave and its dependence on the spin-orbit couplings and on the spin-grating wave-vector. In particular, we determine the optimal parameters for observing the spin-Hall effect with spin gratings. These optimal parameters result from a balance between two competing requirements: 1) To create slowly decaying spin-polarization waves, and 2) To have a strong spin-charge coupling. The first requirement is related to the more general challenge in the field of spintronics of identifying mechanisms allowing for long spin relaxation times. In the presence of disorder, spin-orbit interactions lead to spin relaxation through the Dyakonov-Perel mechanism~\cite{DP}. Recently, it was shown that an  enhanced spin life time can be realized by tuning the spin-orbit coupling so that the Rashba and the linear Dresselhaus couplings become equal~\cite{orenstein}. In this regime, SU(2) spin symmetry is restored, allowing for a long lifetime helical spin density mode termed the ``persistent spin helix''\cite{Bernevig}, provided that the cubic Dresselhaus contribution be minimized~\cite{SG}. However, in the persistent spin helix regime the coupling between the spin and the charge channels vanishes and the spin Hall effect cannot be observed. Hence, the second requirement: the existence of a strong spin-charge coupling.

We consider a two-dimensional electron gas in a III-V type semiconductor quantum well grown along the [001] axis (set as the z axis). In the presence of Rashba~\cite{Rash}, as well as linear and cubic Dresselhaus~\cite{Dress} spin-orbit interactions, the Hamiltonian describing the conduction band electrons is
\begin{equation}
{\cal H}=\frac{{\bf p}^2}{2m}+\mathbf{h}({\bf p})\centerdot\mathbf{\hat{\boldsymbol{\sigma}}}
\end{equation}
where $m$ is the effective mass, $\boldsymbol{\hat{\sigma}}=(\hat{\sigma}_x,\hat{\sigma}_y)$ are Pauli matrices and $\mathbf{h}(\mathbf{p})=(h_x,h_y)$ is the momentum-dependent effective ''magnetic'' field describing the spin-orbit interaction. The Rashba contribution, $\mathbf{h}^R(\mathbf{p})=\alpha v_F(-p_y,p_x)$, with $v_F$ the Fermi velocity,  arises from the inversion asymmetry of the quantum well confining potential, with the coupling constant $\alpha$ measuring the strength of the Rashba spin-orbit coupling relative to the Fermi energy. In addition, the lack of inversion symmetry of the semiconductor crystal gives rise to the Dresselhaus spin-orbit interaction, $\mathbf{h}^{D_1}(\mathbf{p})=\beta_1 v_F(p_x,-p_y)$ and $\mathbf{h}^{D_3}(\mathbf{p})=-4\beta_3 \frac{v_F}{p_F^2}(p_x p_y^2, -p_y p_x^2)$, where $p_F$ is the Fermi momentum and $\beta_1$ and $\beta_3$ are dimensionless coupling constants for the linear and cubic Dresselhaus terms, respectively.
In the presence of disorder, the coupled spin and charge dynamics can be described by a generalized diffusion equation, which in the absence of an external electric field has the form~\cite{BNM,SG} 
\begin{equation}
(\partial_t-\D \nabla^2)\rho_i = (\Gamma^{ij}-P^{ijk}\nabla_k+\mathbf{C}^{ij}\cdot\boldsymbol{\nabla})\rho_j, \label{sce}
\end{equation}
where $\rho_0$ is the charge density and  $\rho_{1,2,3}\equiv\rho_{x,y,z}$ are spin densities. The parameters $\Gamma^{ij}$ describe the Dyakanov-Perel spin relaxation~\cite{DP}, $\D=\tau v_F^2/2$ is the diffusion constant, with $\tau$ the mean scattering time, $P^{ijk}=-P^{jik}$  characterize the precession of the spin polarization and and $\mathbf{C}^{ij}$ describe the coupling between the spin and charge degrees of freedom. In momentum space, the diffusion equation becomes $\left[\delta_{ij}-\hat{\Pi}_{ij}(\omega,\mathbf{k})\right]\rho_j=0$, where  $\hat{\Pi}_{ij}$ have coefficients given by $\Gamma^{ij}$, $P^{ijk}$ and $\mathbf{C}^{ij}$~\cite{SG}. The formal solution of the diffusion equation is $\rho_i(\mathbf{r}, t) =\int d\mathbf{r}^\prime D_{ij}(\mathbf{r},\mathbf{r}^\prime,t)\rho_j(\mathbf{r}^\prime,0)$, where $\rho_i(\mathbf{r},0)$ is the initial spin-charge distribution and $\hat{D}=[\hat{1}-\hat{\Pi}]^{-1}$ is the Green's function of the diffusion equation, or the diffuson. 

The generalization of the spin-charge diffusion formalism developed in Ref. \onlinecite{SG} for the case of a uniform electric field amounts to the formal substitution~\cite{HKCSS}
\begin{equation}
\boldsymbol{\nabla}\rightarrow\boldsymbol{\nabla}+\mu \boldsymbol{E}/2\D,  \label{sub}
\end{equation}
where $E$ is a uniform electric field and $\mu$ is the mobility of the two-dimensional electron gas. Note that, neglecting the spin-charge coupling, this substitution generates the standard drift-diffusion equation for the charge channel, while the description of the spin sector is in agreement with a semi-classical kinetic theory of electron spin transport derived using the Keldysh Green's function formalism~\cite{HKCSS}. The substitution (\ref{sub}) is valid as long as non-linear contributions of order $\mathcal{O}(E^2)$ are small and assuming that the effects of the electron-electron Coulomb interaction can be neglected.  Without loss of generality, we focus on the geometry corresponding to Fig. \ref{Fig1} and  consider a system with an initial out of plane spin-density wave, $\rho_z(r,0)=n_0\mathrm{cos}(qr_+)$, oriented along the $[110]$ direction ($\mathbf{e}_+$) and a weak constant electric field, $\mathbf{E}=E_0\mathbf{e}_-$, oriented along $[1\bar{1}0]$ ($\mathbf{e}_-$). In momentum space, the substitution (\ref{sub}) becomes $\mathbf{k}\rightarrow \mathbf{k}-i\mu \mathbf{E}/\D$ and the inverse of the diffuson is
\begin{equation}
	\hat{1}-\hat{\Pi}(\omega,q)=
	\begin{pmatrix}
		s-1	& i \lambda_- \E 	& \lambda_+ q 	& 0\\
		i \lambda_- \E 	& s-\dg 	& 0 	& -i \gp q\\
		\lambda_+ q 	& 0 	& s+\dg 	& -\gm \E\\
		0 		& i \gp q 	& \gm \E 	& s+1
	\end{pmatrix},
\end{equation}
where $\E=\mu E L_s/2\D$ is a dimensionless measure of the electric field strength and $s=-i\omega(q)+q^2+1$. All lengths are measured in units of spin relaxation length, $L_s = 1/2p_F\Delta$ and times in units of spin relaxation time, $\tau_s = 2\tau/g^2\Delta$, where $\Delta=\left(\alpha^2+(\beta_1-\beta_3)^2+\beta_3^2\right)^{1/2}$ and $g=2v_F p_F\tau$ is a dimensionless conductance. The spin-spin coupling parameters~\cite{SG},  $\gamma_\pm=\gamma_R\pm\gamma_D$,  with $\gamma_R=2\alpha/\Delta $ and $\gamma_D=2(\beta_1-\beta_3)/\Delta $, are independent of the overall strength of the spin-orbit interaction $\Gamma = \left(\alpha^2+\beta_1^2+\beta_3^2\right)^{1/2}$ and lie within  a disc of radius 2. The spin-charge coupling parameters~\cite{SG}, $\lambda_\pm=\lambda_1\pm\lambda_2$, with $\lambda_1=[(3\beta_3-\beta_1)(\alpha^2-\beta_1^2+\beta_3^2)-\beta_1\beta_3^2]/\Delta$ and $\lambda_2=\alpha(\alpha^2-\beta_1^2+6\beta_3^2)/\Delta$,  are quadratic in the spin-orbit interaction strength.

\begin{figure}[tbp]
\begin{center}
\includegraphics[width=0.45\textwidth]{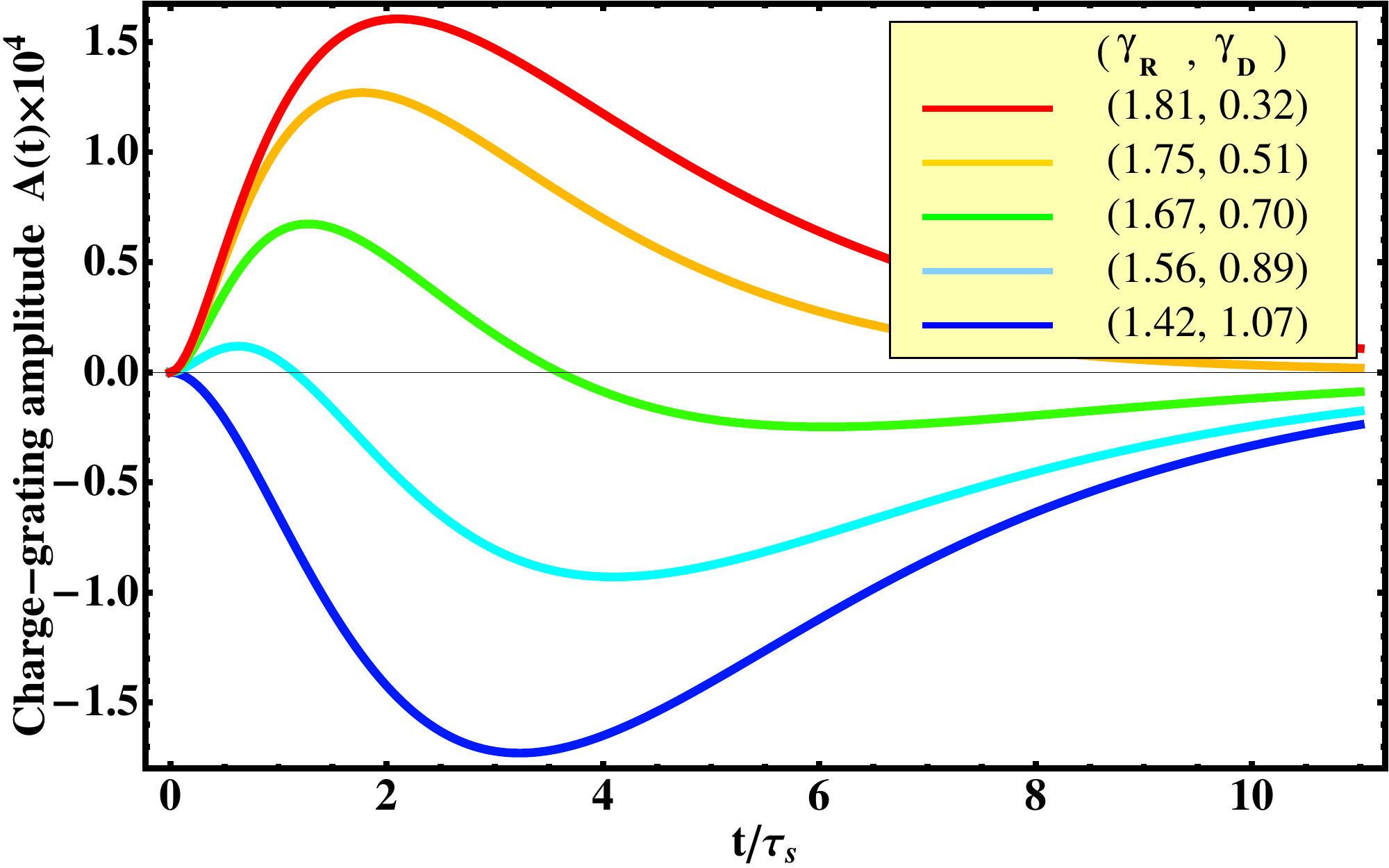}
\end{center}
\caption{(Color online) Time dependence of the induced charge-density wave amplitude A(t) for various values of the dimensionless spin orbit coupling parameters $\pspace$ and for an overall spin-orbit coupling strength $\Gamma=.001$. The wave-vector $\mathbf{q} \parallel \mathbf{e}_+$ has a  fixed value, $q=0.6/L_s$. The amplitude of the induced spin-density wave varies non-monotonically and is characterized by a peak value $A_{max}$ and an exponential decay at large times. \vspace*{-0.25in}} \label{Fig2}
\end{figure}

The induced charge density $\rho_0(\mathbf{r},t)$ is determined by the matrix element $D_{03}=\sum_{l=0}^3A_l(\mathbf{q})e^{-i\omega_l(\mathbf{q})t}$ of the diffuson. Here  $i\omega_l(\mathbf{q})$ are the relaxation modes obtained from the equation $\mbox{det}\left[\hat{1}-\hat{\Pi}(\omega, {\bf q})\right] = 0$ and $A_l(\mathbf{q})$ are momentum-dependent amplitudes.  To order $\mathcal{O}(\E^2)$ the relaxation times are independent of the electric field, while the amplitudes have a linear dependence,  $A_l(\mathbf{q}) = i(\mu E/v_F) q \tilde{A}_l(q)$, with $\tilde{A}_l(q)$ being an even function of momentum and $q=\mathbf{q}\cdot \mathbf{e}_+$. Note that, if one initially generates a charge density profile, the external electric field induces a spin wave with a spatial and time dependence determined by $\hat{D}_{30}=-\hat{D}_{03}$. Hence the present analysis applies to both the direct and the inverse spin Hall effect. Explicitly, an initial out-of-plane spin density wave $\rho_z(r,0)=n_0\mathrm{cos}(qr_+)$ induces a time-dependent charge density wave
\begin{equation}
\rho(\mathbf{r},t)=n_0 \sin(qr_+) \dfrac{\mu E}{v_{f}} \sum_{l=0}^3 q\tilde{A}_l(q)e^{-i\omega_l(q^2)t} \label{cdw}
\end{equation}
Note that the induced charge density wave (CDW) is phase shifted by $\pi/2$ relative to the initial spin-density wave (see Fig. \ref{Fig1}) and has a time dependent amplitude $n_0 (\mu E/v_F) A(t)$ where $A(t)=\sum_{l=0}^3 q\tilde{A}_l(q)e^{-i\omega_l(q^2)t}$. The general behavior of the induced CDW amplitude $A(t)$ is shown in Fig. \ref{Fig2}. At $t=0$ the amplitude of the CDW vanishes, as the system is initially uniform, while  at long times $A(t)$ decays exponentially with a characteristic lifetime $1/(i\omega_l(q))$ given by the lowest frequency relaxation mode. At intermediate times of order $\tau_s$ the CDW amplitude has one maximum and/or one minimum. The largest absolute value defines the peak amplitude $A_{max}$.
The strength of the Rashba and Dresselhaus spin-orbit interaction in GaAs quantum wells can be adjusted by varying the doping asymmetry or the width of the quantum wells. Values in the range of $\alpha=0.5 \times 10^{-3}\div1.5 \times 10^{-3}$ and $\beta_1=1 \times 10^{-3}\div3 \times 10^{-3}$ with $\beta_3=0.3 \times 10^{-3}$ \cite{orenstein} can be experimentally achieved, thus most of region in the vicinity of the boundary of the radius 2 disc in the $\pspace$ parameter space can be probed. Scaling $\alpha$, $\beta_1$ and $\beta_3$ equally will not change the spin-spin couplings $\gamma_R$ or $\gamma_D$, but it will change the spin-charge couplings $\lambda_+$ and $\lambda_-$ which are quadratic in the overall spin-orbit coupling strength $\Gamma$. The amplitudes  $\tilde{A}_l(q)$ depend linearly on $\lambda_+$ and $\lambda_-$ with higher order corrections of order $\lambda_\pm^3$. Thus for experimentally realizable two-dimensional spin-orbit interacting electron systems characterized by $\Gamma\ll 1$, the higher order corrections due to the spin-charge couplings are negligible and the amplitude $A(t)$ is approximately linear in the spin-charge couplings. Since the factor of $q$ in $A(t)$ gives a contribution of $1/\Gamma$, as the wave-vector is measured in units of $1/L_s$, we conclude that the amplitude $A(t)$ of the induced charge-density wave depends linearly on the overall spin-orbit interaction strength $\Gamma$. This proportionality relation holds as long as we express the wave-vector in units of $1/L_s$. Furthermore, we find that the induced CDW amplitude $A(t)$ is independent of the dimensionless conductance g, provided time is measured in units of $1/\tau_s$. Consequently, the bulk manifestation of the spin-Hall effect proposed here, can be enhanced by reducing the carrier density of system, which will increase the ratio between the strength of the spin-orbit interaction and the Fermi energy.
\begin{figure}[tbp]
\begin{center}
\includegraphics[width=0.44\textwidth]{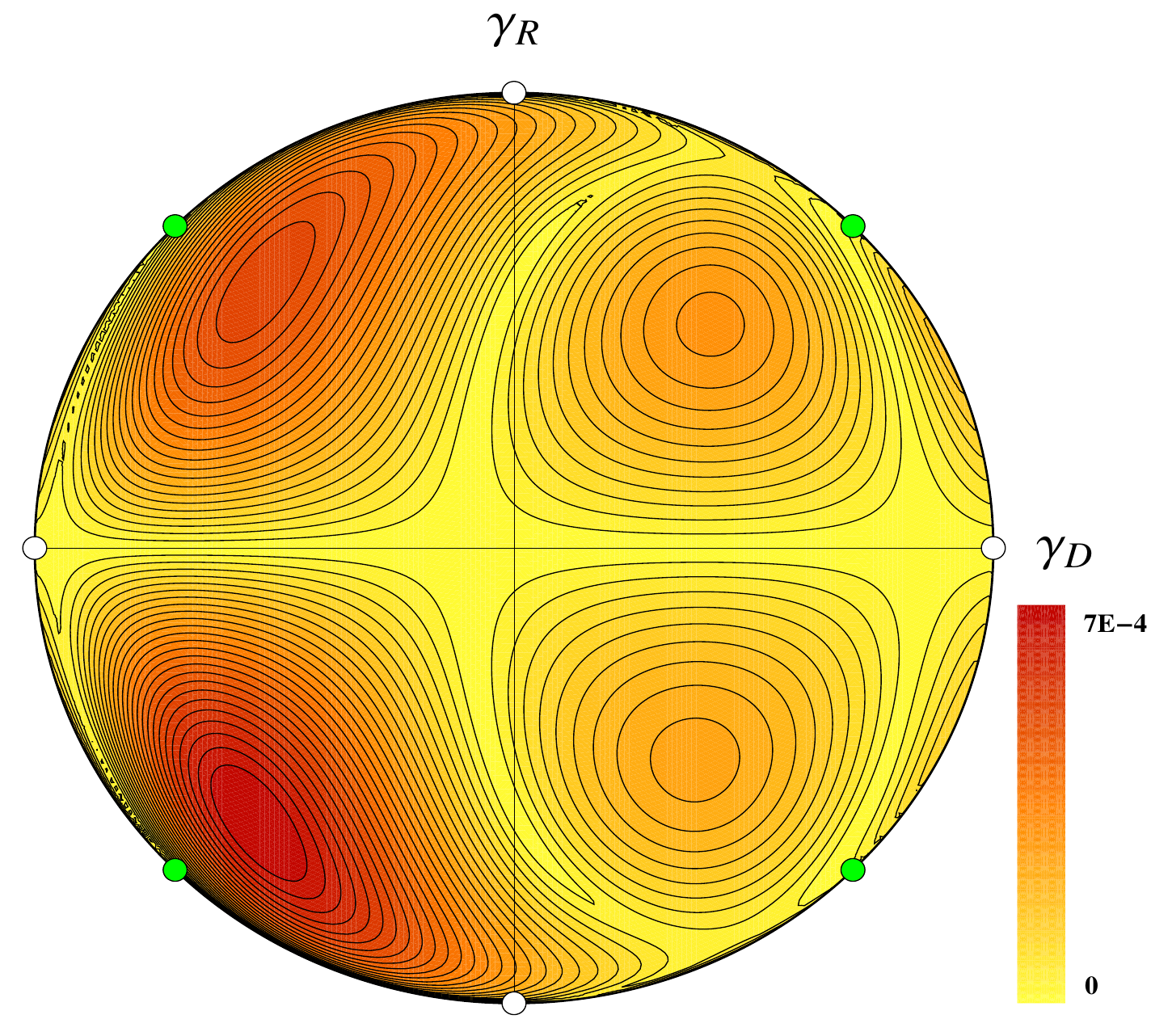}
\end{center}
\caption{(Color online) Dependence of the absolute value of the peak amplitude $A_{max}$ on the spin-orbit parameters $(\gamma_D, \gamma_R)$ for a fixed value of the overall spin-orbit interaction strength, $\Gamma=.001$, and $q=0.7/L_s$.  $A_{max}$ vanishes for pure Dresselhaus spin-orbit coupling, $\gamma_R=0$ (the segment between the horizontal pair of white dots), pure Rashba coupling, $(\gamma_D=0, \gamma_R = \pm2)$ (vertical pair of white dots), and at the symmetry points $(\gamma_D = \pm \sqrt{2}, \gamma_R = \pm \sqrt{2})$ (green dots). The maximum of the peak amplitude corresponds to $(\gamma_D, \gamma_R) = (-1.08, -1.25)$ (inside the lower left quarter of the parameter space, $A_{max}=7.8\times10^{-4}$), while three other local maxima are located at $(\gamma_D, \gamma_R) = (-1, 1.24)$ (upper left quarter, $A_{max}= -5.5\times10^{-4}$), $(0.80, 1.06)$ (upper right, $A_{max}=3.2\times10^{-4}$), and $(0.76, -0.98)$ (lower right, $A_{max}= -2.8\times10^{-4}$). All these maxima involve large relative contributions of the cubic Dresselhaus coupling, $\beta_3/\Gamma = 0.5 \div 0.68$.  \vspace*{-0.15in}} \label{Fig3}
\end{figure}

Next, we study the dependence of the induced charge density wave amplitude on the ratio between various components of the spin-orbit interaction for a fixed value of the overall  spin-orbit coupling strength $\Gamma$. Fig. \ref{Fig3} shows the maximum amplitude of the charge-density wave, $A_{max}$, for the experimentally relevant  spin-orbit coupling strength $\Gamma=0.001$ and wave-vector $q=0.7/L_s$. The peak amplitude vanishes for pure Dresselhaus spin-orbit coupling, $\gamma_R=0$, pure Rashba coupling, $(\gamma_D=0, \gamma_R = \pm2)$, and at the symmetry points $(\gamma_D = \pm \sqrt{2}, \gamma_R = \pm \sqrt{2})$ which support the persistent spin helix modes (see Fig. \ref{Fig3}). This is consistent with previous results showing that, at least in uniform and stationary conditions, the spin Hall conductivity in systems with pure Rashba or pure linear Dresselhaus spin-orbit interaction vanishes~\cite{noSH}. Our analysis reveals the absence of any manifestation of the spin-Hall effect for these types of spin-orbit interactions in non-uniform systems and under time-dependent conditions. The absolute maximum of the peak amplitude, $A_{max}=7.8\times10^{-4}$, is realized for $(\gamma_D, \gamma_R) = (-1.08, -1.25)$. The corresponding original spin-orbit couplings are $(\alpha, \beta_1, \beta_3) = (-7.4, 0.3, 6.7)\times 10^{-4}$. Several other local maxima (minima) can be identified throughout the parameter space (see Fig. \ref{Fig3}). To enhance the peak amplitude of the induced charge profile, one has to consider systems with strong cubic Dresselhaus spin-orbit coupling. This condition is opposite to that required for the realization of the persistent spin helix mode~\cite{SG,orenstein}. Note that the diagram in  Fig. \ref{Fig3} has no particular symmetry, as a result of the nontrivial dependence of the spin-charge coupling parameters $\lambda_{\pm}$ on the spin-orbit couplings. 

\begin{figure}[tbp]
\begin{center}
\includegraphics[width=0.46\textwidth]{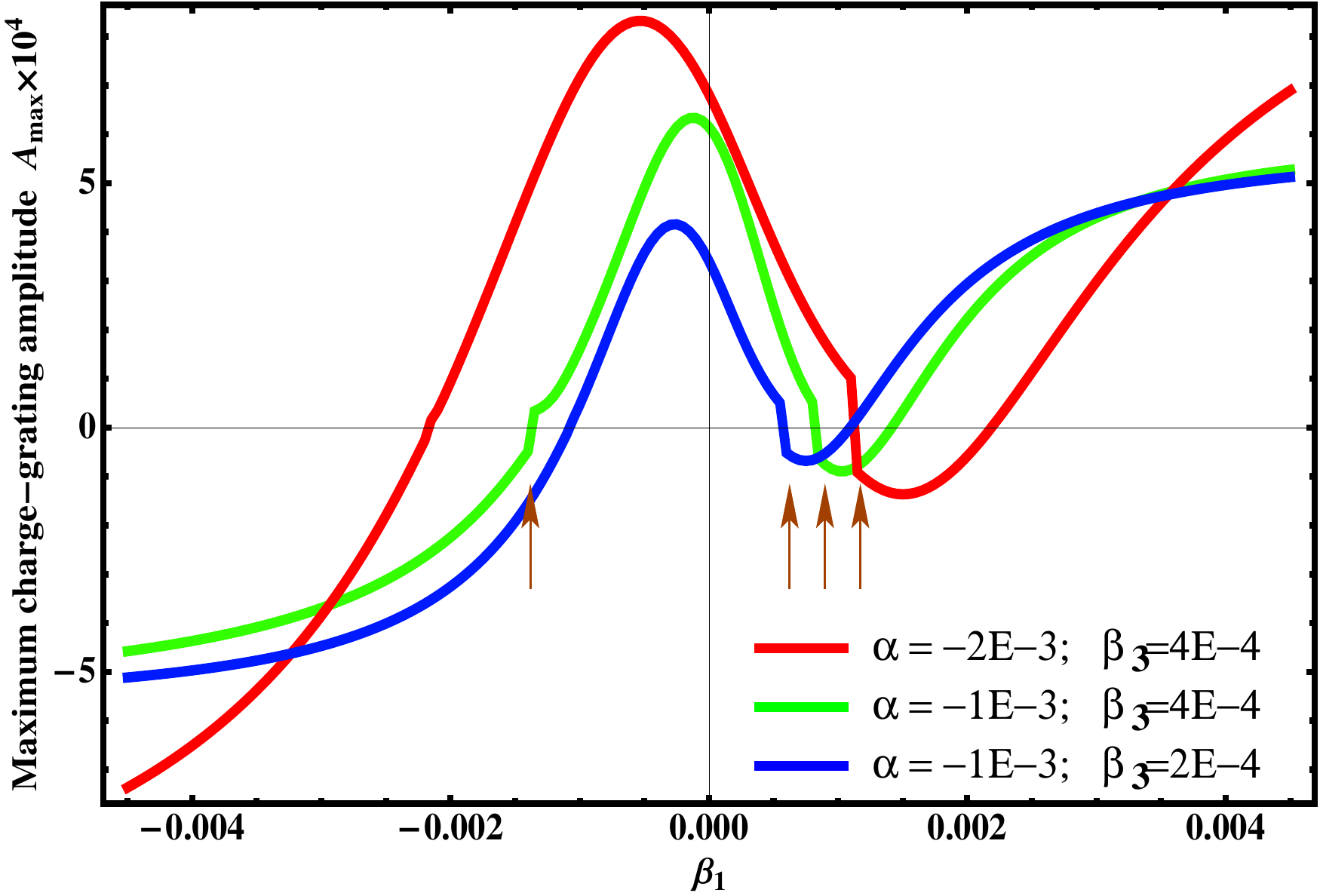}
\end{center}
\caption{(Color online)  Dependence of the peak amplitude on the linear Dresselhaus spin-orbit coupling for various values of the cubic Dresselhaus and Rashba couplings. The arrows mark the values of $\beta_1$  where the $A_{max}$ changes from an absolute minimum of $A(t)$ to an absolute maximum (see also Fig. \ref{Fig2}). \vspace*{-0.2in}} \label{Fig4}
\end{figure}

We consider now the case of a fixed cubic Dresselhaus coupling in the range $\beta_3 = 2\times10^{-4}\div 4\times10^{-4}$, which is experimentally relevant for GaAs quantum wells. The dependence of the peak amplitude on the tunable parameters $\alpha$ and $\beta_1$ is shown in Fig. \ref{Fig4}. We stress that both the absolute value and the sign of the spin-orbit coupling constants are important in determining the strength of the spin-Hall effect. Finally, we note that the peak amplitude also depends on the wave-vector $q$. $A_{max}$ vanishes in the limits $q\rightarrow0$ and $q\rightarrow\infty$ and is maximized in the range $0.5\leq q L_s\leq 0.7$. Increasing the spin-orbit interaction strength enhances the bulk spin-Hall effect, provided it is observed at larger wave-vector values.

For completeness we note that, if the initial spin-density waves have an arbitrary orientation of the q-vector, a charge-density wave is induced even in the absence of an external electric field. However, this wave is in-phase with the initial spin wave. Adding an external electric field perpendicular to the wave-vector induces an additional charge density component characterized by a $\pi/2$ phase shift, as described above, and causes the spin and charge profiles to drift along a direction parallel to the q-vector, i.e., perpendicular to the electric field. The induced charge-density wave has the form
\begin{eqnarray}
\rho({\bf r}, t) &=& n_0\sum_{l=0}^{3} e^{-i\omega_l({\bf q}) t}\left\{a_l({\bf q})\cos\left[{\bf q}\cdot {\bf r} +\left({\bf q}\times\tilde{\bf E}\right)_z\tilde{\Omega}_l({\bf q})t\right]\right.  \nonumber \\
&+&  \left.\left({\bf q}\times\tilde{\bf E}\right)_z\tilde{A}_l({\bf q})\sin\left[{\bf q}\cdot {\bf r} +\left({\bf q}\times\tilde{\bf E}\right)_z\tilde{\Omega}_l({\bf q})t\right]\right\}, \label{IndChar}
\end{eqnarray}
where $a_l({\bf q})$ are the amplitudes of the in-phase charge component and $\omega_l({\bf q})$ are the corresponding frequencies. The electric field induces out-of-phase waves with amplitudes $\tilde{A}_l({\bf q})$ and generates oscillatory components of the relaxation modes proportional to $\tilde{\Omega}_l({\bf q})$. 

In summary, we show that a non-homogeneous spin-orbit interacting system supports bulk manifestations of the spin-Hall effect. We extend the spin-charge diffusion equations to the case of a constant electric field and use this tool to characterize the charge density wave induced by an initial spin density wave that relaxes in the presence of an external electric field perpendicular to the spin-polarization  wave-vector. We find that the induced charge profile is characterized by the same wave-vector as the spin-density wave but has a phase shift of $\pm \pi/2$. The amplitude of the induced charge-density wave varies non-monotonically in time and is characterized by a peak value and an exponential decay at large times. We show how to maximize the effect by tuning the relative strengths of the spin-orbit interactions. Finally, we mention that similar non-homogeneous perturbations may lead to bulk manifestations of the topological quantum spin-Hall effect~\cite{KaneMele,BernZhang} in spin-orbit interacting insulators~\cite{TopBSHE}.

{\bf Acknowledgements:} We thank Joe Orenstein and Jake Koralek for illuminating discussions and a number of valuable suggestions. This work is supported through JQI-PFC; T.S. was supported by LPS-CMTC and US-ONR.

\vspace{-6mm}

\bibliography{SHEwGratings}

\end{document}